\UseRawInputEncoding

\documentclass[]{spie}  

\usepackage{amsmath,amsfonts,amssymb}
\usepackage{graphicx}
\usepackage[colorlinks=true, allcolors=blue]{hyperref}

\title{Diffusion Probabilistic Models for Compressive SAR Imaging}

\author[a]{Odysseas Pappas}
\author[a]{Perla Mayo}
\author[b]{Andrew Austin}
\author[a]{Alin Achim}
\affil[a]{School of Computer Science, University of Bristol, Bristol, UK}
\affil[b]{School of Electrical, Electronic and Mechanical Engineering, University of Bristol, Bristol, UK}

\authorinfo{Corresponding author: Odysseas Pappas, o.pappas@bristol.ac.uk}

\begin{document} 
\maketitle

\begin{abstract}
Compressed sensing Synthetic Aperture Radar (SAR) image formation, formulated as an inverse problem and solved with traditional iterative optimization methods can be very computationally expensive. We investigate the use of denoising diffusion probabilistic models for compressive SAR image reconstruction, where the diffusion model is guided by a poor initial reconstruction from sub-sampled data obtained via standard imaging methods. We present results on real SAR data and compare our compressively sampled diffusion model reconstruction with standard image reconstruction methods utilizing the full data set, demonstrating the potential performance gains in imaging quality.
\end{abstract}

\keywords{Synthetic Aperture Radar, Compressive Sensing, Denoising Diffusion Probabilistic Models}

\section{INTRODUCTION}
\label{sec:intro}  

Synthetic Aperture Radar (SAR) is one of the most commonly employed remote sensing modalities of the modern era, being utilized in a great variety of monitoring tasks due to its ability to produce highly detailed images of the Earth regardless of the day-night cycle and weather conditions. SAR is a coherent imaging radar system (typically airborne/spaceborne) that transmits frequency-modulated signals to a scene on the ground and records the reflected radar echoes while in flight, achieving impressive spatial resolution due to its long synthetic aperture. 

As with all coherent radar systems, the collected reflected radar echoes, often referred to as phase history data, must be processed to form an image suitable for visual interpretation. Several classical methods exist for the formation of SAR images from phase history data \cite{Franceschetti1999}, each offering distinct advantages and disadvantages in terms of speed, computational cost and image quality. The sheer volume of the generated phase history data, especially for modern high-resolution SAR systems, imposes severe costs in terms of data storage and transfer as well as processing time. There is therefore significant interest in the formation of SAR images from subsampled phase history data, with compressed sensing (CS) methods for SAR imaging gathering a lot of research interest. 

CS \cite{Donoho2006, Candes2008} is a mathematical framework that allows for the reconstruction of signals from a parsimonious set of measurements (weighted linear combinations of samples) far smaller than what the Nyquist theorem would define as the minimum total of samples required. Applications of CS in SAR\cite{Zhang2022} allow for the reconstruction of a SAR image from a sub-sampled set of phase history data, introducing obvious advantages in terms of data acquisition speed and storage requirements while also allowing for the treatment of erroneously incomplete data. Algorithms for CS SAR image formation often employ an inverse problem formulation combined with sparse regularization methods\cite{Cetin2014, Uḡur2012, Leier2014}. The design of a suitable signal model and measurement matrix for various SAR imaging modes can however be a non-trivial task, and this limits the practicability of such methods. Additionally, due to the iterative nature of the solvers employed in such inverse problem formulation, they can be extremely computationally intensive, especially for high-resolution imagery.  

In this paper we examine an alternative approach taking inspiration from CS SAR but utilizing generative image models such as Denoising Diffusion Probabilistic Models (DDPMs) instead of classical CS reconstruction theory. DDPMs rely on the modelling of a denoising process to generate images from noise samples, with the capability of conditioning the generated image  on a class label or on an input signal. This conditioning input signal can be a low resolution copy or a poor quality reconstruction of the desired image, with the DDPM effectively achieving super-resolution or image enhancement respectively. We explore the use of a CS measurement matrix via which a sub-sampled phase history can be obtained, with this sub-sampled phase history in turn formed into an image using classical imaging methods. While the resulting image is expectedly of very poor quality, we demonstrate how it can serve as conditioning input to a DDPM which in turn can produce an image of comparable quality to what would have been obtained from classical methods operating on the fully sampled phase history. 

The paper is organized as follows: Section 2 provides some additional information on CS and discusses the design of an appropriate measurement matrix. Section 3 discusses DDPMs and their use in our proposed method. Section 4 presents experimental results of our CS SAR DDPM imaging method while concluding remarks are provided in Section 5. 

\section{MEASUREMENT MATRICES FOR CS SAR IMAGING}

 CS \cite{Donoho2006, Candes2008} is a novel sampling framework that allows for the reconstruction of sparse signals from a subsampled set of measurements. CS theory posits that the signal of interest $x$ is sparsely representable in some domain, i.e. there exists a $K$-sparse signal $s$ that satisfies

\begin{equation}
\label{eq:sparsex}
x = \Psi s,     x \in \Re^N
\end{equation}
for some sparsifying transform $\Psi$. CS then provides us with a framework for the reconstruction of $x$ from a set of measurements $y$ obtained via

\begin{equation}
\label{eq:y}
y = C x = C \Psi s,     y \in \Re^M
\end{equation}
where $C$ is the CS measurement matrix and $M<<N$. Typical CS reconstruction methods rely on an inverse problem formulation whereby a minimization problem

\begin{equation}
\label{eq:smin}
\hat{s} = \arg \min ||s||_1
\end{equation}
is solved such that $C\Psi s = y$.

\begin{figure} [t]
   \begin{center}
   \begin{tabular}{c} 
   \includegraphics[height=4.5cm]{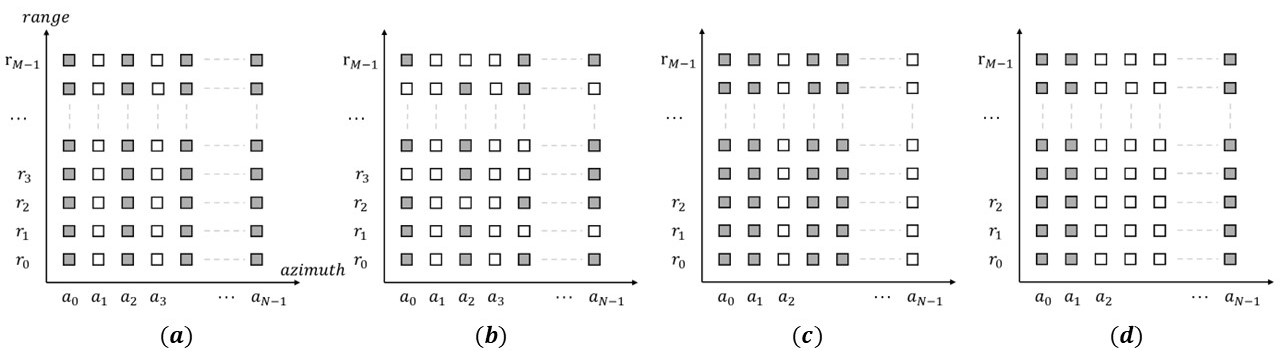}
   \end{tabular}
   \end{center}
   \caption[subsampl] 
   { \label{fig:subsamp} 
    Subsampling patterns for a SAR phase history matrix containing $M$x$N$ samples. (a) Regular subsampling (at 50\%) along the azimuth. (b) Regular subsampling across azimuth with random subsampling across range. (c) Random subsampling across azimuth. (d) Aperture gap subsampling.}
\end{figure} 

Central to any CS method is the design of a suitable measurement matrix $C$, as it must satisfy two properties, namely the restricted isometry property (RIP) and incoherence (with respect to $\Psi$). Random $C$ matrices can satisfy these properties, however their use can be impractical in many applications and therefore the design of application-specific CS measurement matrices holds significant research interest. SAR-specific CS measurement matrices have been a topic of discussion in the literature \cite{Leier2014, Uḡur2012, Kelly2012a}. A pattern easily realizable in practice is that of regular undersampling along the azimuth direction, as a simple modification to the A/D converter at the SAR groundstation can allow for sampling at lower rates along the azimuth, in effect discarding entire aperture positions (phase history matrix columns) at regular intervals. 

There is some debate as to whether regular subsampling across the azimuth is sufficient for CS, as it may not satisfy the restricted isometry property \cite{Uḡur2012}. Introducing random dropping of samples along the range direction (on the aperture positions that are not outright discarded) can help alleviate such concerns, though it necessitates much more complex modifications to the SAR hardware \cite{Leier2014}. Random sampling in azimuth alone is also feasible, as is sampling with continuous gaps in the total aperture \cite{Kelly2012a}.Figure \ref{fig:subsamp} demonstrates these subsampling patterns. For the experiments discussed in this paper, we have opted to follow the regular azimuth subsampling pattern shown in Fig. \ref{fig:subsamp}(a) as it is easily realizable in practice. Note RIP adherence is not a major concern since we do not follow conventional CS reconstruction methods but rather rely on DDPM for reconstruction.

\begin{figure} [t]
   \begin{center}
   \begin{tabular}{c} 
   \includegraphics[height=4cm]{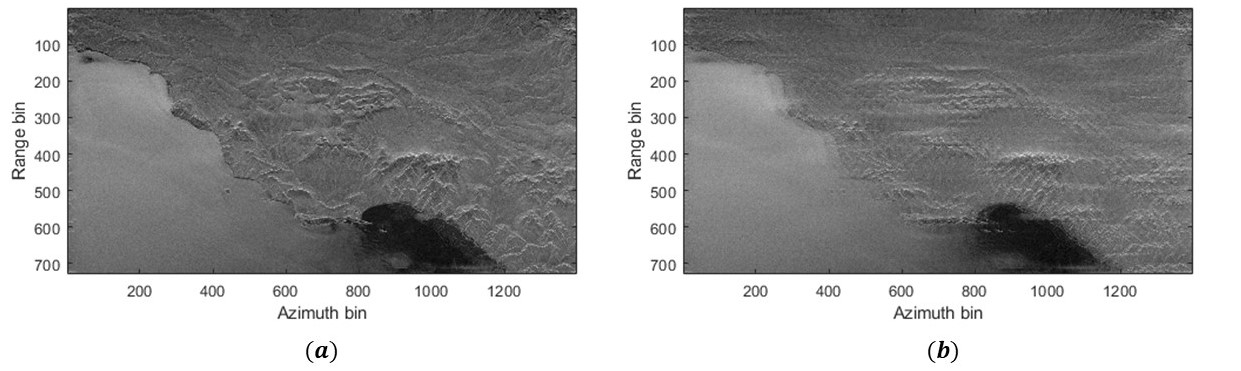}
   \end{tabular}
   \end{center}
   \caption[poorq] 
   { \label{fig:poorq} 
    Effect of phase history subsampling on classical reconstruction methods. (a) ERS-1 image reconstructed from fully sampled phase history data. (b) ERS-1 image reconstructed from phase history data subsampled by a factor of 2 along the azimuth direction. Both reconstructions via the range migration algorithm. Images have been multi-looked with 20 azimuth looks and 4 range looks.}
\end{figure}

Conventional imaging algorithms can be applied to such subsampled phase history data but expectedly fall short of creating a proper SAR image. An example of this effect can be seen in Fig. \ref{fig:poorq}. Phase history data have been extracted from a Level-0 ERS-1 SAR product and formed into an image using the range migration algorithm, producing a clean SAR image of good quality as shown in Fig. \ref{fig:poorq}(a). For Fig. \ref{fig:poorq}(b) the same history data have been subsampled along the azimuth direction by a factor of two; the dimensions of the phase history matrix have been kept constant, with ``discarded" aperture positions (columns) being recorded as zero values (which can be corrupted by noise to simulate the instrument's noise floor). The subsampled phase history has been processed via the same range migration algorithm, with all radar parameters (e.g. the pulse repetition frequency) kept the same. The produced image suffers from very noticeable blurring and especially aliasing artifacts oriented along the azimuth, attributed to the periodic pattern of the subsampling mask. Note that different subsampling patterns would introduce different deteriorations to the image.

\section{DENOISING DIFFUSION PROBABILISTIC MODELS FOR CS SAR}
Denoising Diffusion Probabilistic Models (DDPMs), originally proposed by Ho et al. \cite{Ho2020}, rely on the modeling of a denoising process for synthetic data generation and have in recent years evolved into one of the most promising families of generative AI models. During training, the image is iteratively corrupted by increasing levels of noise (as defined by a noise schedule) to the point where it completely deteriorates into noise. During this forward (diffusion) process the model is trained to predict how the image progressively deteriorates. Application of the reverse (backward) process effectively amounts to a denoising process via which an image can be iteratively structured from noise. 

In more detail, assuming $p(x_0)$ is the probability density distribution of the training data, with the $0$ index indicating the original, uncorrupted index, given such an uncorrupted input image $x_0$, it becomes progressively corrupted via the addition of Gaussian noise in versions $x_1,x_2,...,x_T$ obtained according to the Markovian process 

\begin{equation}
\label{eq:ddpmfwd}
p(x_t|x_{t-1}) = \mathcal{N} (x_t; \sqrt{1-\beta_t} \cdot x_{t-1}, \beta_t \cdot \textbf{I}), \forall t\in \{1,...,T\},
\end{equation}
where $T$ is the total number of steps in the noise variance schedule (i.e. diffusion steps), $\beta_1,...,\beta_T$ are hyperparameters representing the variance schedule across diffusion steps, $\textbf{I}$ is the identity matrix (of equal dimensions to the input image $x_0$) and $\mathcal{N}(x;\mu,\sigma)$ is the normal distribution of mean $\mu$ and variance $\sigma$ that produces $x$.

To sample at any arbitrary timestep $t$ we rewrite equation 
\ref{eq:ddpmfwd} using $\alpha_t=1-\beta_t$ and $\overline\alpha_t=\Pi_{s=1}^t\alpha_s$ as:

\begin{equation}
\label{eq:ddpmfwd2}
p(x_t|x_0) = \mathcal{N} (x_t; \sqrt{\alpha_t} \cdot x_0, (1 - \overline{\alpha}) \cdot \textbf{I})
\end{equation}
And therefore, given that the variances of the noise schedule are pre-set and not learnable, the sampling of $x_t$ can be denoted as:

\begin{equation}
\label{eq:ddpmfwd3}
x_t = \sqrt{\overline{\alpha_t}} \cdot x_0 + \sqrt{1-\overline{\alpha_t}\epsilon}, \epsilon~\mathcal{N}(0,\textbf{I})
\end{equation}

In applying the reverse process, a DDPM can generate new samples from $p(x_0)$ by beginning with a random Gaussian noise sample and iteratively forming it into a structured image by following the operation:

\begin{equation}
\label{eq:ddpmbwd}
p(x_{t-1}|x_t) = \mathcal{N}(x_{t-1};\mu(x_t,t),\Sigma_\theta(x_t,t)),
\end{equation}
where $\Sigma_\theta(x_t,t)=\sigma^2_t\textbf{I}$ is a fixed variance.

DDPMs are capable of conditioning the generated image to belong to a certain class, provided class labels were provided alongside the training data. More recently, the ability to condition the model not on a class label but rather on an input image have further unlocked the potential of DDPMs to be used for a number of tasks such as computational image formation, in-painting and super-resolution \cite{Croitoru2023}. 

The performance of DDPMs has been iteratively augmented in recent years. The first major improvement came with the introduction of Improved DDPM (IDDPM) \cite{Nichol2021}. IDDPM brings in a number of updates in the network structure and training process to improve log-likelihood performance of DDPMs to a level comparable to that of Generative Adversarial Networks (GANs). This was followed by Guided DDPM (GDDPM) \cite{Dhariwal2021}, incorporating further improvements that greatly enhanced the ability to produce outputs conditioned on an input image. In the context of SAR, DDPMs have been applied to a number of tasks including synthetic SAR image generation \cite{Crump2024, Qosja2024} (whether unconditional or class-conditioned). Wang et al. \cite{Wang2024} have recently presented a SAR imaging algorithm that utilises DDPMs as a prior term under a statistical plug-and-play imaging framework; note however that this work does not touch on CS for SAR.

Here we propose the use of DDPMs for SAR imaging conditioned on a poor-quality image produced via conventional methods from CS-sampled data. A GDDPM model akin to that used for super-resolution applications is jointly trained over matched pairs of good quality images (produced from fully sampled phase histories) and poor quality images (produced by CS-sampled phase histories). DDPMs have been employed in similar (albeit non-SAR related) fashion in DOLCE \cite{Liu2022}, where the authors posit Limited Angle Computed Tomography image reconstruction via a DDPM conditioned on a poor-quality subsampled reconstruction, interleaved with a data consistency mechanism.

DDPMs utilises U-Net as the backbone neural network, as do many similar diffusion models. Scaling such a network to natively handle images of a size in line with typical SAR image products is not practically feasible, as it would pose enormous demands in terms of computational power and data storage. The processing of large images can be addressed via a number of methods, including multi-scale/hierarchical methods and latent space diffusion \cite{Rombach2022}, with an obvious alternative also being the adoption of a patching strategy \cite{Ozdenizci2023}. 

We opt for the later here and design our network to operate on image patches. The data loader extracts a single patch from each training image, with is location randomly set; this allowing the network to see multiple possible patches from each image as the training process iterates. For sampling, the conditioning input image is segmented into overlapping tiles and these are processed independently through the network and then restructured into the original image dimensions, with an optional histogram normalization step to minimize blocking artifacts in the re-assembled image.


\section{Experimental Results}

\begin{figure} [t]
   \begin{center}
   \begin{tabular}{c} 
   \includegraphics[height=7.5cm]{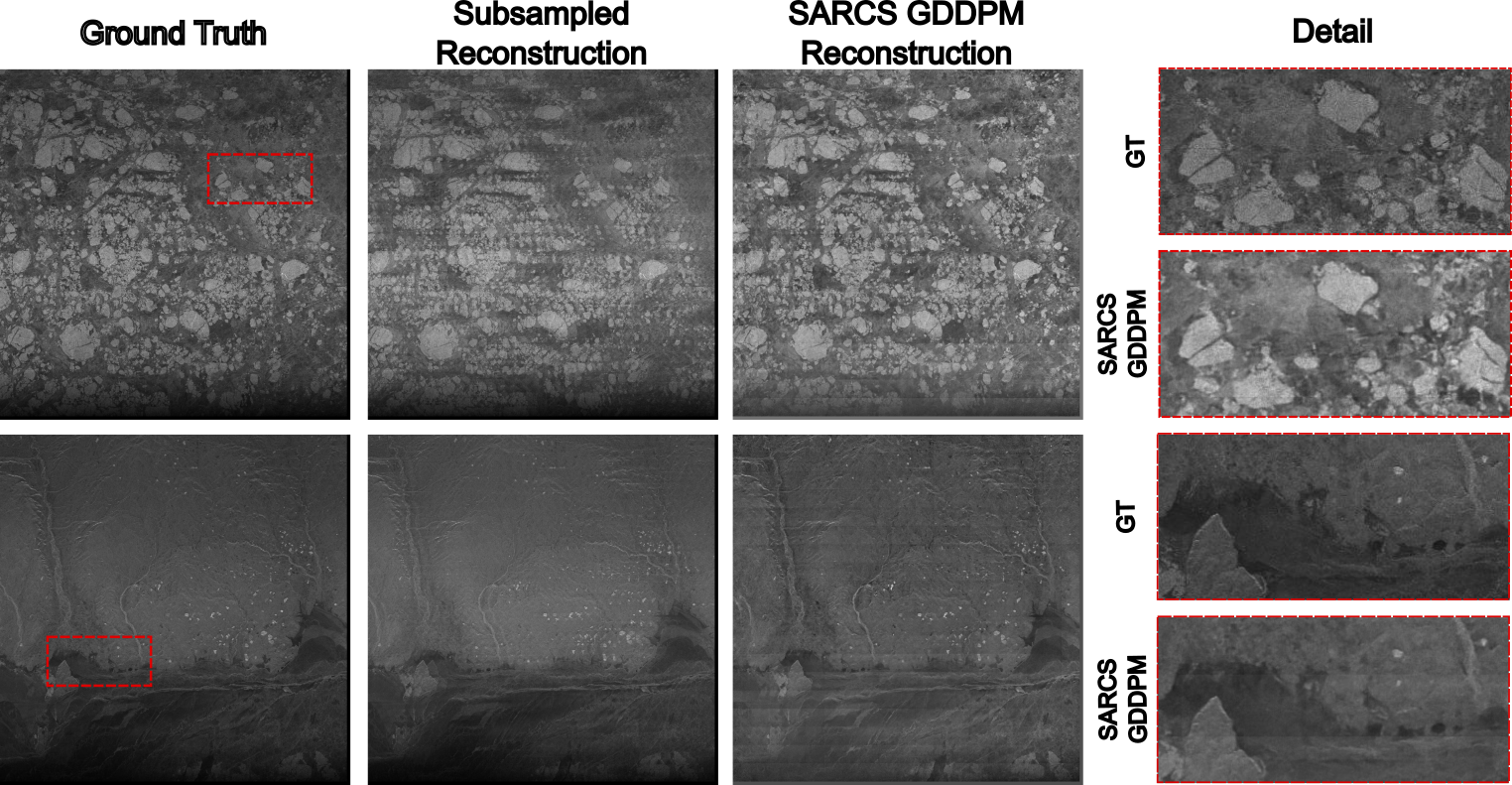}
   \end{tabular}
   \end{center}
   \caption[resultl] 
   { \label{fig:result} 
    Sample experimental results. Ground truth images are constructed from fully sampled phase histories via range migration. Subsampled reconstructions (serving as conditional input to SARCS GDDPM) constructed from subsampled phase histories (50\% across azimuth). Red square designates detail area.}
\end{figure} 

For our experiments we have compiled a proprietary dataset of ERS SAR products. Our training dataset consists of 500 ERS-1 Level 0 products acquired during January 1996 over North America. Phase history data have been decoded and subsequently formed into images using the standard range migration algorithm \cite{Franceschetti1999}. Additionally, the original phase history data were subsampled at a ratio of 50$\%$ regularly across the azimuth, with noise floor values substituting the signal values in the original matrix columns corresponding to these non-sampled aperture positions. The subsampled phase histories were also processed via the standard range migration algorithm, producing a dataset of very poor quality, aliased reconstructions. In both cases the produced images are then multi-looked with 20 looks in the azimuth and 4 in the range direction, and cropped to 1228x1228 pixels (and zero-padded as needed depending on patch size) . These matched pairs of fully-sampled and poor-quality CS-sampled reconstructions serve as the training data for our algorithm.

While the ERS mission is one of the oldest SAR missions and the images produced by it are not representative of the capabilities of modern SAR platforms, their use was motivated by the need to access, decode and process large volumes of Level 0 raw data. As low-quality conditioning inputs must be generated from sub-sampled phase history data the use of Level 1 formed image data was not possible, and few other missions besides ERS-1 provide large quantities of usable raw Level 0 data.

Figure \ref{fig:result} shows reconstruction results using GDDPM conditioned on CS-sampled reconstructions. These have been produced with a GDDPM model trained over 80,000 iterations, with 1000 diffusion steps, a learning rate of 1e-4 and a linear noise schedule. The network is sized to handle image patches that are 256x256 pixels and the conditioning image has been tiled with a stride of 64. The U-Net backbone has 192 base feature channels, 2 residual blocks per resolution level and attention layers at resolutions of 32x32 and below. The test images used are derived from ERS-1 products not included in the 500 training images.

It can be seen, especially in the detailed view of the two provided examples, that SARCS GDDPM is capable of alleviating the blurring and aliasing effects prominent in the conditioning CS-sampled input to a great extent, producing an image comparable to that obtained from fully sampled data. Some tiling artifacts and inconsistencies in contrast can be seen in the SARCS GDDPM results and can be addressed either at post-processing or by enforcing closer adherence to the original condition contrast during the GDDPM sampling process. 

\section{Conclusions and Future Work}
Our preliminary results demonstrate that GDDPMS can be used to produce satisfactory SAR images from undersampled phase histories which would have otherwise produced images suffering from severe aliasing and blurring effects. We intend to investigate further the effect of various CS measurement matrices as well as network architectures, and to develop more elaborate post-processing algorithms for de-blocking and contrast correction, while also investigating the embedding of statistical conditioning on the generated images in future research.

\acknowledgments 
 
This work was funded by DASA under the Defence Rapid Impact Open Call.

\bibliography{cssar} 
\bibliographystyle{spiebib} 

\end{document}